\documentstyle[12pt,epsf]{article}
\def\Journal#1#2#3#4{{#1} {\bf #2}, #3 (#4)}

\def\NPB{{\em Nucl. Phys.} B}
\def\PLB{{\em Phys. Lett.}  B}
\def\PRL{\em Phys. Rev. Lett.}
\def\PRD{{\em Phys. Rev.} D}

\begin{document}
\begin{titlepage}
\hspace{12cm}
YERPHI-1588(9)
\vspace{1cm}
\begin{center}
{\Large Complete bremsstrahlung corrections to the forward-backward asymmetries
in $b\rightarrow X_s\ell^+\ell^-$}
\end{center}
\vspace{3cm}
\begin{center}
{\large
H. M. Asatrian,  H. H. Asatryan, A. Hovhannisyan, V. Poghosyan}\\
\vspace{1cm}
Yerevan Physics Institute, 2 Alikhanyan Br., 375036, Yerevan, Armenia
\end{center}
\vspace{3cm}
In a recent paper we presented a calculation of NNLL virtual corrections
to the forward-backward asymmetries in $b\rightarrow X_s\ell^+\ell^-$ decay.
That result does not include bremsstrahlung corrections
which are free from infrared and collinear singularities.
In the present paper we include the remaining ${\cal O}(\alpha_s)$ bremsstrahlung
corrections to  the forward-backward asymmetries in $b\rightarrow X_s\ell^+\ell^-$
decay. The numerical effect of the calculated contributions is found to be below 1\%.
\end{titlepage}
\section{Introduction}
\hspace{0.4cm}
Rare B-decays are known to provide a unique source of  information about the physics at the scales of several
hundred GeV. In the standard model (SM) all these decays proceed through loop diagrams and
are  suppressed. Thus in  the extensions of  the SM the contributions from the
'new' sources of flavor-violation can be comparable to or even larger than the SM contribution.
Therefore experimental information on rare decays can be used to test the SM at the
one-loop level or to put constraints on its extensions.

During the last decade the experimental and theoretical efforts were concentrated on the
$b\to s\gamma$ mediated decays. Available experimental data already provides stringent
constraints on certain extensions of the SM.
In 2002, also the exclusive  $B\rightarrow K_s\mu^+\mu^-$ decay mode was measured by the BELLE collaboration
\cite{Abe:2001dh}. This measurement was confirmed soon by the BABAR Collaboration \cite{Aubert:2002pj}.
Recently, also the first measurement of the branching ratio in the inclusive decay
$B\rightarrow X_s\ell^+\ell^-$ has been reported by the BELLE Collaboration \cite{Kaneko:2002mr}.
The results of these first measurments are compatible with SM predictions though more statistics
will be needed for more decisive conclusions.
It is expected that precise measurement of kinematical distributions
for the $B\rightarrow X_s\ell^+\ell^-$ decay combined with data on
$B\rightarrow X_s\gamma$ will significantly tighten the constraints on the extensions of the standard model
\cite{Ali:2002jg}.

>From the theoretical point of view the description of $B\rightarrow X_s\ell^+\ell^-$ decay is problematic
because of the  long-distance contributions from intermediate $c\bar{c}$ resonant states.
When the invariant mass $\sqrt{s}$ of lepton pair is close to the mass of resonance, only
model dependent predictions for these long distance contributions are available today.
However for the region $0.05<\hat{s}=s/m_b^2<0.25$ the nonperturbative effects
are estimated to be below 10\% \cite{Falk:1994dh}-\cite{Krueger:1996} and the differential decay rate
for $B\rightarrow X_s\ell^+\ell^-$ can be well approximated by HQET corrected short distance
contribution.

The next-to-leading logarithmic (NLL) calculation  for $B\rightarrow X_s\ell^+\ell^-$
has been performed quite long ago in \cite{Misiak:1993bc,Buras:1995dj}.
However those results are known to suffer from a relatively
large ($\pm16\%$) dependence on the matching scale $\mu_W$.
The NNLL corrections to the Wilson coefficients eliminate the matching scale dependence to a large extent
\cite{Bobeth:2000}, but leave a $\pm13\%$-dependence on the renormalization
scale $\mu_b$, which is of order ${\cal O}(m_b)$. To further improve the theoretical prediction,
${\cal O}(\alpha_s)$ virtual and bremsstrahlung corrections have been calculated
\cite{Asatrian:2001de}-\cite{Asatryan:2002iy}.
As a result the renormalization scale dependence reduced by a factor of 2.

It is well known that the measurement of the forward-backward asymmetries for decay
 $B\rightarrow X_s\ell^+\ell^-$ can be used, in combination with the measurement
 of $B\rightarrow X_s \gamma$,
to perform a so-called model independent test of the standard model \cite{Ali:2002jg}, \cite{goto},\cite{ali1}.
The double differential decay
width $d\Gamma/[d\hat{s}dcos{\theta}]$ and the forward-backward
asymmetries have recently been calculated with NNLL precision \cite{Asatrian:2002va},\cite{Ghinculov:2002pe}.
Those calculations included one-loop virtual ${\cal O}(\alpha_s)$
corrections associated with the
operators $O_7$, $O_9$ and $O_{10}$ as well as the corresponding
bremsstrahlung corrections that are necessary to cancel the infrared and mass singularities.
It was found that NNLL corrections drastically reduce the renormalization
scale dependence of forward-backward asymmetries.
In the present paper we complete the calculation of the NNLL calculation for the
forward-backward asymmetries, presenting the full results for the
bremsstrahlung corrections associated with the operators $O_1$,
$O_2$, $O_8$  which were omitted in
\cite{Asatrian:2002va},\cite{Ghinculov:2002pe}.

The paper is organized as follows. In section 2 we briefly describe the
theoretical framework. In section 3  the analytical results  for the forward-backward
asymmetries in $b\rightarrow X_s\ell^+\ell^-$ decay are presented.
In section 4  we discuss the technical details of the calculations and give
phenomenological analysis for the forward-backward asymmetries  in $b\rightarrow X_s\ell^+\ell^-$ decay .

\section{The Theoretical Framework}
\hspace{0.4cm}
The most efficient tool for studies on weak B meson decays is the effective Hamiltonian technique. For the specific
 channels $b\rightarrow sl^+l^-$ $(l=\mu ,e)$, the effective Hamiltonian is of the form\\
\begin{equation}
{\mathcal
H}_{eff}=-\frac{4G_F}{\sqrt2}V_{ts}^{*}V_{tb}\sum_{i=1}^{10}C_iO_i
\end{equation}
where we have omitted the contribution proportional to the small CKM factor $V_{us}^{*}V_{ub}$.
The dimension six effective operators
can be chosen as
\begin{eqnarray}
\nonumber && O_1=(\bar s_L \gamma_\mu T^a c_L)(\bar c_L \gamma^\mu T^a b_L),   \,\ \,\  \,\ \,\  \,\ \,\
O_2=(\bar s_L \gamma_\mu c_L)(\bar c_L \gamma^\mu b_L),\\
\nonumber && O_3=(\bar s_L \gamma_\mu b_L)\sum_q (\bar q \gamma^\mu q), \,\ \,\  \,\ \,\  \,\ \,\  \,\ \,\ \,\ \,\
 O_4=(\bar s_L \gamma_\mu T^a b_L)\sum_q (\bar q \gamma^\mu T^a q),\\
\nonumber && O_5=(\bar s_L \gamma_\mu\gamma_\nu\gamma_\rho b_L)\sum_q (\bar q
\gamma^\mu\gamma_\nu\gamma_\rho q), \,\ \,\
O_6=(\bar s_L \gamma_\mu\gamma_\nu\gamma_\rho T^a b_L)\sum_q (\bar q \gamma^
\mu\gamma_\nu\gamma_\rho T^a q),\\
&& O_7=\frac {e}{g_s^2}m_b(\bar s_L \sigma^{\mu\nu}b_R)F_{\mu\nu},  \,\ \,\  \,\ \,\  \,\ \,\  \,\ \,\ \,\
O_8=\frac {1}{g_s}m_b(\bar s_L\sigma^{\mu\nu}b_R)F_{\mu\nu}, \\
\nonumber && O_9=\frac {e^2}{g_s^2}(\bar s_L \gamma_\mu b_L)(\bar\ell \gamma_\mu \ell),  \,\ \,\  \,\ \,\  \,\ \,\  \,\
 \,\  \,\ \,\
O_{10}=\frac {e^2}{g_s^2}(\bar s_L \gamma_\mu b_L)(\bar\ell \gamma_\mu \gamma_5 \ell)
\end{eqnarray}
where the subscripts $L$ and $R$ refer to left- and right- handed
components of the fermion fields. In the following it is convenient
to use the related operators $\tilde{O}_7,...,\tilde{O}_{10}$,
defined according to
\begin{equation}
\tilde{O}_j=\frac {\alpha_s}{4\pi}O_j  \,\ \,\ \,\ \,\ \,\
(j=7,...,10)
\end{equation}
with the corresponding Wilson coefficients
\begin{equation}
\tilde{C}_j=\frac {4\pi} {\alpha_s} C_j \,\ \,\ \,\ \,\ \,\
(j=7,...,10).
\end{equation}
We refer to \cite{Bobeth:2000} and \cite{Asatryan:2001zw} for details.

\section{NNLL results for the forward-backward asymmetries}
\hspace{0.4cm}
We start introducing the forward-backward asymmetries. Often the so-called
normalized and un-normalized forward-backward asymmetries are considered which are
defined by
\begin{equation}
\bar A_{FB}(\hat s)=\frac {\int_{-1}^1 [d^2\Gamma (b\rightarrow
X_s\ell^+\ell^-)/d\hat sdz]sgn(z)dz}{\int_{-1}^1 [d^2\Gamma (b\rightarrow
X_s\ell^+\ell^-)/d\hat sdz]dz}
\end{equation}
and
\begin{equation}
A_{FB}(\hat s)=\frac {\int_{-1}^1 [d^2\Gamma (b\rightarrow
X_s\ell^+\ell^-)/d\hat sdz]sgn(z)dz}{\Gamma (B\rightarrow X_ce\bar
\nu_e)}BR_{sl}
\end{equation}
respectively.

The double differential decay witdh can be written in the form \cite{Asatrian:2002va}:
\begin{eqnarray}
\label{dgammadsdz}
\nonumber && \frac{d^2\Gamma (b\rightarrow
X_s\ell^+\ell^-)}{d\hat{s} dz}=
\left(\frac{\alpha_{e.m.}}{4\pi}\right)^2 \frac{G^2_Fm_{b,pole}^5
|V_{ts}^{*}V_{tb}|^2}{48\pi^3}(1-\hat s)^2 \\
\nonumber && \times \left\{{\frac{3}{4}[(1-z^2)+\hat
s(1+z^2)](|\tilde C_9^{eff}|^2+|\tilde C_{10}^{eff}|^2) \left (1+
\frac {2\alpha_s}{\pi}f_{99}(\hat s,z) \right)}\right.\\
\nonumber && \left.+ \frac {3}{\hat s}[(1+z^2)+\hat s (1-z^2)]
|\tilde C_7^{eff}|^2 \left (1+\frac {2\alpha_s}{\pi}f_{77} (\hat
s,z) \right) - 3\hat s z Re(\tilde C_9^{eff}
\tilde C_{10}^{eff\ast}) \left (1+\frac {2\alpha_s}{\pi}f_{910}(\hat s) \right)\right. \\
&& \left.+6Re(\tilde C_7^{eff}\tilde C_9^{eff\ast}) \left (1+\frac
{2\alpha_s}{\pi}f_{79}(\hat s,z) \right)-
6zRe(\tilde C_7^{eff}\tilde C_{10}^{eff\ast}) \left (1+\frac {2\alpha_s}{\pi}f_{710}(\hat s) \right)\right\}\\
\nonumber && + \frac{d^2\Delta \Gamma (b\rightarrow
X_s\ell^+\ell^-)}{d\hat{s} dz}
\end{eqnarray}
where $z=cos(\theta)$ and the effective Wilson coefficients
$\tilde C_7^{eff}$, $\tilde C_9^{eff}$ and $\tilde C_{10}^{eff}$ are given by
\begin{eqnarray}
\label{effcoeff}
\nonumber
&& \tilde C_7^{eff}=A_7-\frac
{\alpha_s(\mu)}{4\pi}[C_1^{(0)}F_1^{(7)}(\hat
s)+C_2^{(0)}F_2^{(7)}(\hat s)+A_8^{(0)}F_8^{(7)}(\hat s)],\\
\nonumber && \tilde C_9^{eff}=A_9+T_9h(\hat{m}_c^2,\hat{s})+U_9h(1,\hat{s})+W_9h(0,\hat{s}) \\
&& \,\ \,\ \,\ \,\ \,\ \,\ \,\
-\frac{\alpha_s(\mu)}{4\pi}[C_1^{(0)}F_1^{(9)}(\hat{s})+C_2^{(0)}F_2^{(9)}(\hat
s)+
A_8^{(0)}F_8^{(9)}(\hat s)],\\
&&\tilde C_{10}^{eff}=A_{10},
\nonumber
\end{eqnarray}
with $\hat{m}_c^2=m_c^2/m_b^2$.
The quantities $C_1^{(0)}$, $C_2^{(0)}$, $A_7$, $A_8^{(0)}$, $A_9$, $A_{10}$,
$T_9$, $U_9$ and $W_9$ are Wilson
coefficients or linear
combinations thereof. Their analytic expressions can be found in
\cite{Asatrian:2001de}-\cite{Asatrian:2002va}. In Eq.~(\ref{dgammadsdz}) the term
$\frac{d^2\Delta \Gamma (b\rightarrow
X_s\ell^+\ell^-)}{d\hat{s} dz}$ is due to the finite bremsstrahlung corrections that have not
been considered in \cite{Asatrian:2002va}. In this paper our goal will be to investigate the impact of that
term on  forward-backward asymmetries. The general
distribution (\ref{dgammadsdz}) will be investigated elsewhere \cite{prep}.

In the numerator, both asymmetries involve the same integral that can be expressed as follows:
\begin{eqnarray}
\label{numerator}
\nonumber &&  \int_{-1}^1\frac {d^2\Gamma (b\rightarrow
X_s\ell^+\ell^-)}{d\hat s dz}sgn(z)dz
 =  \left(\frac{\alpha_{e.m.}}{4\pi}\right)^2\frac{G^2_Fm_{b,pole}^5
|V_{ts}^{*}V_{tb}|^2}{48\pi^3}(1-\hat s)^2 \\
\nonumber && \times \left[- 3\hat s Re\left(\tilde C_9^{eff}
\tilde C_{10}^{eff\ast}\right) \left (1+\frac
{2\alpha_s}{\pi}f_{910}(\hat s) \right)-6Re\left(\tilde
C_7^{eff}\tilde C_{10}^{eff\ast}\right) \left
(1+\frac {2\alpha_s}{\pi}f_{710}(\hat s) \right)\right. \\
&& \left.+
\frac{1}{3}\frac {2\alpha_s}{\pi}\left(\,Re\left(\tilde
C_8^{0,eff}\tilde C_{10}^{eff\ast}\right) \,t_{810}(\hat s)
 +Re\left(\left(C_2^{0}-\frac{1}{6}C_1^{0}\right) \tilde C_{10}^{eff\ast}\right)
 \,t_{210}(\hat s)\right)\right]
\end{eqnarray}
The functions $f_{710}$ and $f_{910}$ have been calculated in \cite{Asatrian:2002va}
while the terms proportional to $t_{210}$ and $t_{810}$ arise from the interference of
the matrix elements of the operators $O_1, O_2, O_8$ with $O_{10}$ and are
calculated in the present paper. Their explicit expressions read:
\begin{eqnarray}
\nonumber  t_{810}&=& \frac{1}{6(1-\hat{s})^2}((3((1-\sqrt{\hat{s}})^2(23-6\sqrt{\hat{s}}-\hat{s})+
4(1-\hat{s})(7+\hat{s})\ln(1+\sqrt{\hat{s}}) \\
\nonumber &+&2s(1+s-\ln(\hat{s}))\ln(\hat{s})))+2(-3\pi^2(1+2\hat{s})+
6(3-\hat{s})\hat{s} \ln(2-\sqrt{\hat{s}}) \\
\nonumber &
-&36(1+2\hat{s})Li_2[-\sqrt{\hat{s}}]-\sqrt{\frac{\hat{s}}{4-\hat{s}}}(6(2(-3+\hat{s})\hat{s}
\arctan\left[\frac{2+\sqrt{\hat{s}}}{\sqrt{4-\hat{s}}}\right] \\
\nonumber &+&2\pi \ln(2-\sqrt{\hat{s}})-\arctan\left[\sqrt{\frac{4-\hat{s}}{\hat{s}}}\right]((-3+\hat{s})\hat{s}+
4\ln(2-\sqrt{\hat{s}})) \\
\nonumber &-&\arctan\left[\frac{\sqrt{\hat{s}}\sqrt{4-\hat{s}}}{2-
\hat{s}}\right]((-3+\hat{s})\hat{s}-\ln(\hat{s}))+4Re(i
Li_2[\frac{(-2+i \sqrt{4-\hat{s}}+\sqrt{sh})\sqrt{\hat{s}}}
{i \sqrt{4-\hat{s}}-\sqrt{\hat{s}}}] )\\
 &-&2Re(i Li_2[\frac{i}{2}\sqrt{4-\hat{s}}(1-\hat{s})
\sqrt{\hat{s}}+\frac{(3-\hat{s})\hat{s}}{2}])))))
\end{eqnarray}
\begin{eqnarray}
\nonumber
t_{210}&=&\int^{1}_{\hat{s}}\frac{-\hat{s}}{(\hat{s}-w)(1-\hat{s})^2}\{(4(1-\hat{\hat{s}})(1+w)-\frac{2\sqrt{(\hat{s}-w^2)^2}
(w(3+w)-\hat{s}(1-w))}{w^2}\\
\nonumber
&+&(2+5w+2w^2+\hat{s}(3+4w))\ln\left[\frac{\hat{s}+w^2+\sqrt{(\hat{s}-w^2)^2}}{2w}\right]
-\frac{(\hat{s}-w)}{\hat{s}\sqrt{(1+w)^2-4\hat{s}}}\\
\nonumber &
\times &(w(2-w)-\hat{s}(6-5w))
(\ln(1+w-\hat{s}(3-w)+(1-\hat{s})\sqrt{(1+w)^2-4\hat{s}})\\
\nonumber &-&\ln(\hat{s}(1-3w)+w^2(1+w)+\sqrt{(s-w^2)^2}\sqrt{(1+w)^2-4\hat{s}})))\bar{\Delta}_{23}\\
\nonumber &-&(2(1-\hat{s})(1+2w)-\frac{2\sqrt{(\hat{s}-w^2)^2}(w(2+w)-\hat{s}(1-w))}{w^2}\\
\nonumber &+&2(\hat{s}(1+2w)+w(2+w))\ln\left[\frac{\hat{s}+w^2+\sqrt{(\hat{s}-w^2)^2}}{2w}\right]\\
\nonumber &
+&\frac{4(1-w)(\hat{s}-w)}{\sqrt{(1+w)^2-4\hat{s}}}(\ln(1+w-\hat{s}(3-w)+(1-\hat{s})\sqrt{(1+w)^2-4\hat{s}})\\
 &-&\ln(\hat{s}(1-3w)+w^2(1+w)+\sqrt{(\hat{s}-w^2)^2}\sqrt{(1+w)^2-4\hat{s}}))){\bar{\Delta}_{27}}\}dw
\end{eqnarray}
where
\begin{eqnarray}
\nonumber
&& \bar{\Delta}_{i_{23}}=-2+\frac{4}{w-\hat{s}}\left[zG_{-1}\left(\frac{\hat{s}}{z}\right)-zG_{-1}\left(\frac{w}{z}\right)
-\frac{\hat{s}}{2}G_0\left(\frac{\hat{s}}{z}\right)+\frac{\hat{s}}{2}G_0\left(\frac{w}{z}\right)\right],\\
&& \bar{\Delta}_{i_{27}}=2\left[G_0\left(\frac{\hat{s}}{z}\right)-G_0\left(\frac{w}{z}\right)\right],
\end{eqnarray}
The functions $G_{-1}(t)$ and $G_{0}(t)$ can be found in
\cite{Asatryan:2002iy} (Eqn. (30), (31)).\\
The functions $t_{210}$ and $t_{810}$ are plotted in Fig.1.

\section{The details of the calculation and numerical results}
\hspace{0.4cm}
As in \cite{Asatrian:2002va} we will work in the rest frame of the lepton pair. The corresponding
formulae have been derived in \cite{Asatrian:2002va}.
For the 3-momenta of
b-quark ($\vec{p}_b$), $\ell^+$ ($\vec{l}_2$) and gluon ($\vec{r}$)
in that frame of reference we have:
\begin{eqnarray}
\nonumber && \vec{p}_b=(|\vec{p}_b|,0,0),\,\,\,
\vec{l}_2=(\frac{\sqrt{s}}{2}cos\theta,\frac{\sqrt{s}}{2}sin\theta,0),\\
\nonumber && \vec{r}=(E_rcos\theta_1,E_rsin\theta_1cos\theta_2,E_rsin\theta_1sin\theta_2),
\end{eqnarray}
where
\begin{eqnarray}
\nonumber cos\theta_1=\frac{2E_b\sqrt{s}-2E_r\sqrt{s}+2E_rE_b-s-m_b^2+m_s^2}{2E_r|\vec{p}_b|},
\end{eqnarray}
$E_b$ is the energy of the $b$ quark and $E_r$ is the energy of gluon.
Then the formula for double differential decay width reduces to
\begin{eqnarray}\label{phaseInt}
\frac{d^2\Gamma (b\rightarrow sg\ell^+\ell^-)}{d\hat{s}dz}=
\frac{m_b\hat s}{(2\pi)^62^5}\times \int
\bar{|M|^2}(1-{z_2}^2)^{-1/2}dz_2dE_rdE_b
\end{eqnarray}
where $\bar{|M|^2}$ is  the squared matrix element,
summed and averaged over spins and colors of the particles in the final and initial states,
respectively, $\hat{s}=s/m_b^2$ and $z=cos\theta$, $z_2=cos\theta_2$.
The boundaries for the integration variables are
\begin{eqnarray}
\nonumber &&
E^{min}_r=\frac{{m_b}^2+s-2E_b\sqrt{s}-{m_s}^2}{2(E_b+|\vec{p_b}|-\sqrt{s})}\leq
E_r\leq\frac{{m_b}^2+s-2E_b\sqrt{s}-{m_s}^2}{2(E_b-|\vec{p_b}|-\sqrt{s})}=E^{max}_r\\
&& m_b\leq E_b \leq \frac{m_b^2+s-m_s^2}{2\sqrt{s}}, \,\ \,\ \,\
-1\leq z_2\leq1
\end{eqnarray}

As we have mentioned above the forward-backward asymmetries arise only from interference of $O_{10}$
matrix element with $O_i, i=1,..,9$\footnote{As in our previous papers we will systematically ignore the
${\cal O}(\alpha_s)$ contributions coming from matrix elements of operators  $O_i, i=3,4,5,6$ as their Wilson
coefficients are very small.}. The infrared infinite bremsstrahlung corrections coming from interference of $O_{10}$
with $O_{7}$ and $O_{9}$ (along with the corresponding virtual corrections) have been taken into account by
introducing functions $f_{710}$ and $f_{910}$ which have been calculated in our previous paper.
The remaining bremsstrahlung corrections are infrared safe and are coming from the interference
of $O_8$ and $O_1$,  $O_2$ with $O_{10}$ (see Fig.2 and Fig.3 for the contributing Feinman diagrams).

The calculation of $O_8$ and $O_{10}$ interference term is relatively easy and can be performed
analytically. The calculation for $O_{1,2}$ and $O_{10}$ interference terms is more complicated
and will be described in some detailes below.

The bremsstrahlung corrections involve the matrix elements associated with the two
diagrams  in Fig. 3. Their sum, $\bar{J}_{\alpha \beta}$, is given by
\begin{eqnarray}
    \bar{J}_{\alpha\beta} &=& \frac{e\,g_s\,Q_u}{16\,\pi^2}
    \left[
        E(\alpha,\beta,r) \, \bar\Delta i_5 +
        E(\alpha,\beta,q) \, \bar\Delta i_6 -
        E(\beta,r,q)\frac{r_{\alpha}}{qr} \, \bar\Delta i_{23}
    \right.\nonumber\\
        &-& \left.E(\alpha,r,q)\frac{q_{\beta}}{qr} \, \bar\Delta i_{26}
        - E(\beta,r,q)\frac{q_{\alpha}}{qr} \, \bar\Delta i_{27}
    \right]
    L \, \frac{\lambda}{2} \, ,
\end{eqnarray}
where $q$ and $r$ denote the momenta of the virtual photon and of the gluon, respectively.
The index $\alpha$ will be contracted with the photon propagator, whereas $\beta$ will
contracted with the polarization vector $\epsilon^\beta(r)$
of the gluon \cite{Asatryan:2001zw}. The matrix
$E(\alpha,\beta,r)$ is defined as
\begin{equation}
    E(\alpha,\beta,r) = \frac{1}{2} (\gamma_{\alpha}\gamma_{\beta}\hat{r} -\hat{r}\gamma_{\beta}\gamma_{\alpha}).
\end{equation}
Due to Ward identities, the quantities $\bar\Delta i_k$ are not independent of one another. Namely,
we have
\begin{equation}
    \bar\Delta i_5 = \bar\Delta i_{23} + \frac{q^2}{qr} \bar\Delta i_{27} \, ; \quad
    \bar\Delta i_6 = \bar\Delta i_{26}\, .
\end{equation}
As in addition $\bar\Delta i_{26} = - \bar\Delta i_{23}$, the bremsstrahlung matrix elements depend on $\bar\Delta
i_{23}$ and $\bar\Delta i_{27}$, only. In $d=4$ dimensions we find
\begin{eqnarray}
    \label{eq:deltaik}
    \bar\Delta i_{23} = 8\, (qr) \int_0^1\! dx\,dy\, \frac{x\, y (1-y)^2}{C} \, ,
    \bar\Delta i_{27} = 8\, (qr) \int_0^1\! dx\,dy\, \frac{y\, (1-y)^2}{C} \, ,
\end{eqnarray}
where
\begin{equation}
    C = m_c^2 - 2\, x\, y (1-y) (qr) - q^2\, y\, (1-y) - i\,\delta.
\end{equation}

It is easy to notice that dependence of the matrix element $M$ on $z_2$ is polynomial
so the integral over $z_2$ in Eq.(\ref{phaseInt}) is straightforward. To deal with the remaining two
dimensional integrals over $\hat{E}_r = E_r/m_b$ and $\hat{E}_b = E_b/m_b$
it is useful to introduce a new integration variable $w$ instead of $\hat{E}_r$ defined by
$w=2\sqrt{\hat{s}}E_r+\hat{s}$ .The integration limits are then given by
\begin{equation}
    \hat{E}_b \in \left[ \frac{\hat{s}+w^2}{2w\sqrt{\hat{s}}},  \frac{1+\hat{s}}{2\sqrt{\hat{s}}}\right]
\quad and \quad w \in [\hat{s},1].
\end{equation}
For a fixed value of $\hat{s}$, the quantities $\bar\Delta i_{23}$ and $\bar\Delta i_{27}$
depend only on the scalar product
$qr$, which  is given by $(w-\hat{s})\,m_b^2/2$. The integration over $\hat{E}_b$ then turns
out to be of rational kind and can be performed analytically. The remaining integration over $w$, however, is more
complicated and is done numerically.

We now proceed to the investigation of the numerical impact of the finite bremsstrahlung corrections
on the forward-backward asymmetries.
In Fig. 4 we show the contribution of the finite bremsstrahlung corrections to the unnormalized
forward-backward asymmetry $\Delta A_{FB}(\hat{s})$.
Dashed-dot lines show the contribution from interference of matrix elements
of the operators $O_8$ and $O_{10}$  ($\mu$ =2.5 GeV (uppermost curve), 5 GeV (middle curve),
and 10 GeV  (lower curve)), dashed lines show the contribution from
interference of matrix elements of the operators $O_{1,2}$ and $O_{10}$  ($\mu$ =2.5 GeV (lower curve),
5 GeV (middle curve),  and 10 GeV (uppermost curve) and
), solid  lines show the sum of previous two terms ($\mu$ =2.5 GeV
(lower curve), 5 GeV (middle curve),  and 10 GeV (uppermost curve)), $m_c/m_b$=0.29.
We see that contributions coming from $O_{1,2}-O_{10}$ and $O_{8}-O_{10}$
interference terms partially cancel each other. For that reason the contribution
of new terms to the forward backward asymmetries is numerically small.
In Fig. 5 we combine the new corrections with the previous results for unnormalized forward-backward
asymmetry. The solid lines show $A_{FB}(\hat{s})$ including new corrections for three
values of renormalization scale ($\mu$=2.5,5 and 10 GeV) and $\hat{m_c}=0.29$. The dashed lines represent the
corresponding results  without finite bremsstrahlung corrections..
In Fig. 6 we give the corresponding curves for normalized   forward-backward asymmetry
$\bar{A}_{FB}(\hat{s})$.
The new contribution is numerically small:
at $\hat{s} =0$ we find
$A_{FB}^{NNLL}(0)=2.304\pm 0.109$
while without finite bremsstrahlung corrections we have
$A_{FB}^{NNLL}(0)=2.292\pm 0.111$. For the position $\hat{s}_0$ where forward-backward
asymmetry is zero we find
$\hat{s}_0^{NNLL}=0.1620\pm 0.0016$
while without finite bremsstrahlung corrections we have
$\hat{s}_0^{NNLL}=0.1615\pm 0.0015$.

It is necessary to mention that for some extensions of the standard model (see for instance, \cite{Kagan1998})
with  opposite sign   (and larger values) of $\tilde C_8^{eff}$ the contribution of the finite bremsstrahlung
corrections to the forward-backward asymmetries can be sizable.

To conclude, we have calculated the finite bremsstrahlung corrections to the
forward-backward asymmetries for $b\rightarrow X_s\ell^+\ell^-$ decay. We found that
the numerical impact of the new corrections on the forward-backward asymmetries in the standard model
is less than 1\%, but for some extensions of the standard model it can be larger.\\

\begin{center}
ACKNOWLEDGMENTS\\
\end{center}
\hspace{0.4cm}
We thank C. Greub  for useful discussions.
The work was partially supported by NFSAT-PH 095-02 (CRDF 12050) and SCOPES programs.

\vspace{3cm}
\def\lb{\raisebox{0.5mm}{\big(}}
\def\rb{\raisebox{0.5mm}{\big)}}
\begin{table}[hbt]
\caption{\label{table1} Coefficients appearing Eq.~(\ref{effcoeff}) for $\mu=2.5$~GeV,
$\mu=5$~GeV and $\mu=10$~GeV. For $\alpha_s(\mu)$ (in the $\overline{\mbox{MS}}$ scheme) we used the two-loop
expression with five flavors and $\alpha_s(m_Z)=0.119$. The entries correspond to the pole top quark mass
$m_t=174$~GeV. The superscript (0) refers to lowest order quantities.}

    \begin{center}
    \begin{tabular*}{\textwidth}{l@{\hspace*{1.5cm}}c@{\hspace*{1.5cm}}c@{\hspace*{1.5cm}}c}
        \hline\hline
        $\mu$                                & $ 2.5$ GeV          & $ 5$ GeV           & $ 10$ GeV           \\
        \hline
        $\alpha_s                          $ & $ 0.2675           $ & $ 0.2150          $ & $ 0.1800           $ \\
        $C_1^{(0)}                         $ & $ -0.6960          $ & $ -0.4860         $ & $-0.3257           $ \\
        $C_2^{(0)}                         $ & $ 1.0458           $ & $ 1.0235          $ & $ 1.0109           $ \\
        $\lb A_7^{(0)},~A_7^{(1)}\rb       $ & $ (-0.3605,~0.0322) $ & $(-0.3207,~0.0185) $ & $ (-0.2873,~0.0089) $ \\
        $A_8^{(0)}                         $ & $ -0.1688          $ & $ -0.1535         $ & $ -0.1403          $ \\
        $\lb A_9^{(0)},~A_9^{(1)}\rb       $ & $ (4.2407,~-0.1700) $ & $(4.1283,~0.0130)  $ & $ (4.1311,~0.1549)  $ \\
        $\lb T_9^{(0)},~T_9^{(1)}\rb       $ & $ (0.1148,~0.2791)  $ & $(0.3744,~0.2512)  $ & $ (0.5763,~0.2307)  $ \\
        $\lb U_9^{(0)},~U_9^{(1)}\rb       $ & $ (0.0455,~0.0228)  $ & $(0.0326,~0.0153)  $ & $ (0.0224,~0.0105)  $ \\
        $\lb W_{9}^{(0)},~W_{9}^{(1)}\rb   $ & $ (0.0440,~0.0162)  $ & $(0.0321,~0.0117)  $ & $ (0.0223,~0.0085)  $ \\
        $\lb A_{10}^{(0)},~A_{10}^{(1)}\rb $ & $ (-4.3731,~0.1349) $ & $(-4.3731,~0.1349) $ & $ (-4.3731,~0.1349) $ \\
        \hline\hline
   \end{tabular*}
    \end{center}
\end{table}

\newpage
\begin{figure}
\begin{center}
\epsfxsize=18cm
\epsfysize=18cm
\leavevmode
\vspace{15cm}
\mbox{\hskip 5cm}
\epsfbox[10 7 673 377]{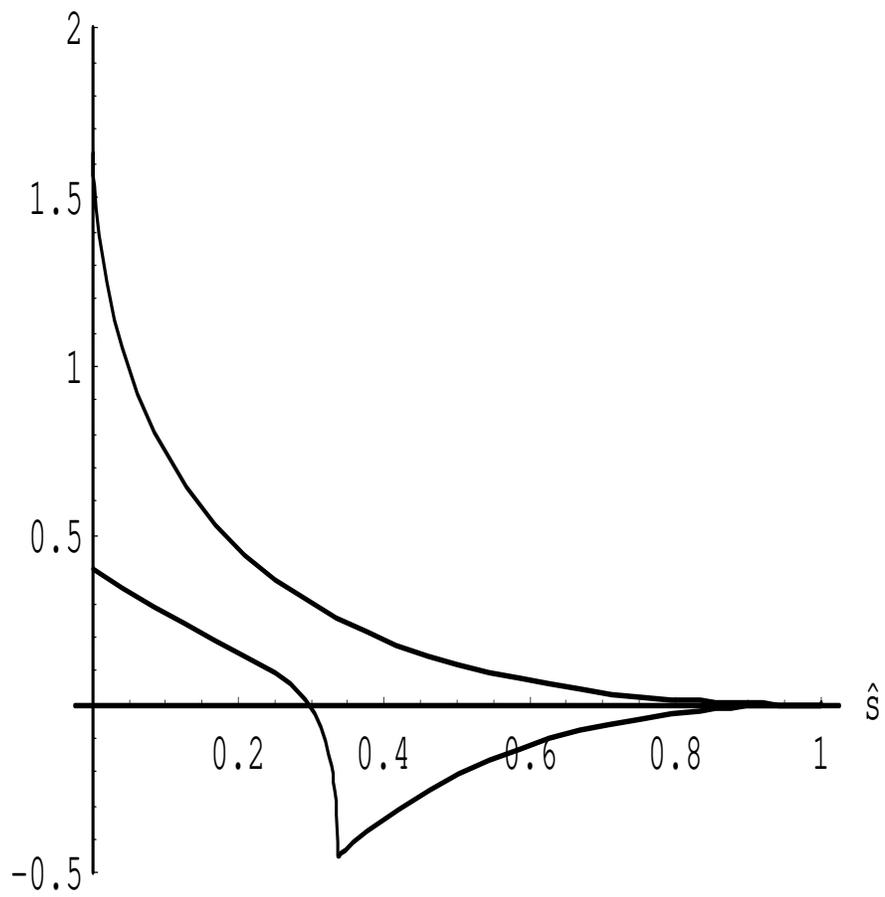}
\end{center}
\vspace{-16cm}
\caption[]{Functions $t_{810}$ (upper curve) and $t_{210}$. }
\end{figure}
\newpage
\begin{figure}
\begin{center}
\epsfxsize=20cm
\epsfysize=10cm
\leavevmode
\vspace{-3cm}
\mbox{\hskip 10cm}
\epsfbox[10 7 673 377]{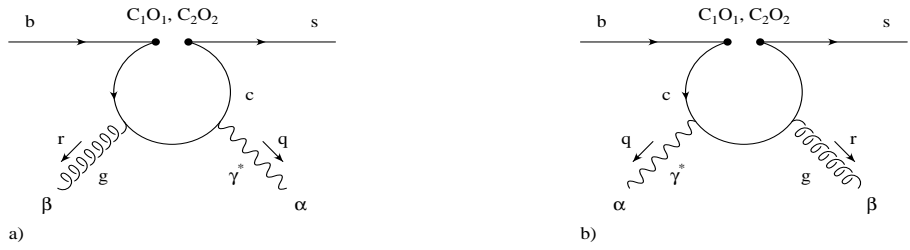}
\end{center}
\vspace{-3cm}
\hspace{-10cm}
\caption[]{
Bremsstrahlung diagrams induced by  $O_1$ and $O_2$.}
\end{figure}
\begin{figure}
\begin{center}
\epsfxsize=18cm
\epsfysize=12cm
\leavevmode
\vspace{-1cm}
\mbox{\hskip -0.4in}
\epsfbox[10 7 673 377]{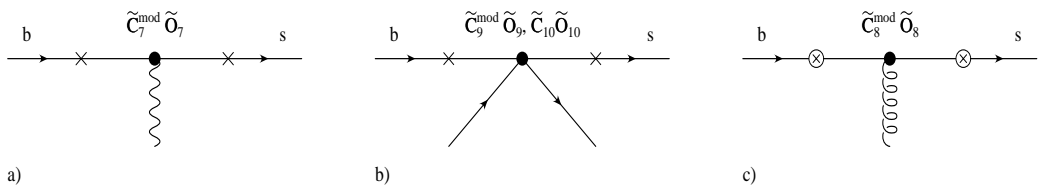}
\end{center}
\vspace{-5cm}
\caption[]{
Bremsstrahlung diagrams induced by  $O_7$ and $O_8$, $O_9$, $O_{10}$.}
\end{figure}
\newpage
\begin{figure}
\begin{center}
\epsfxsize=18cm
\epsfysize=18cm
\leavevmode
\vspace{18cm}
\mbox{\hskip 5cm}
\epsfbox[10 7 673 377]{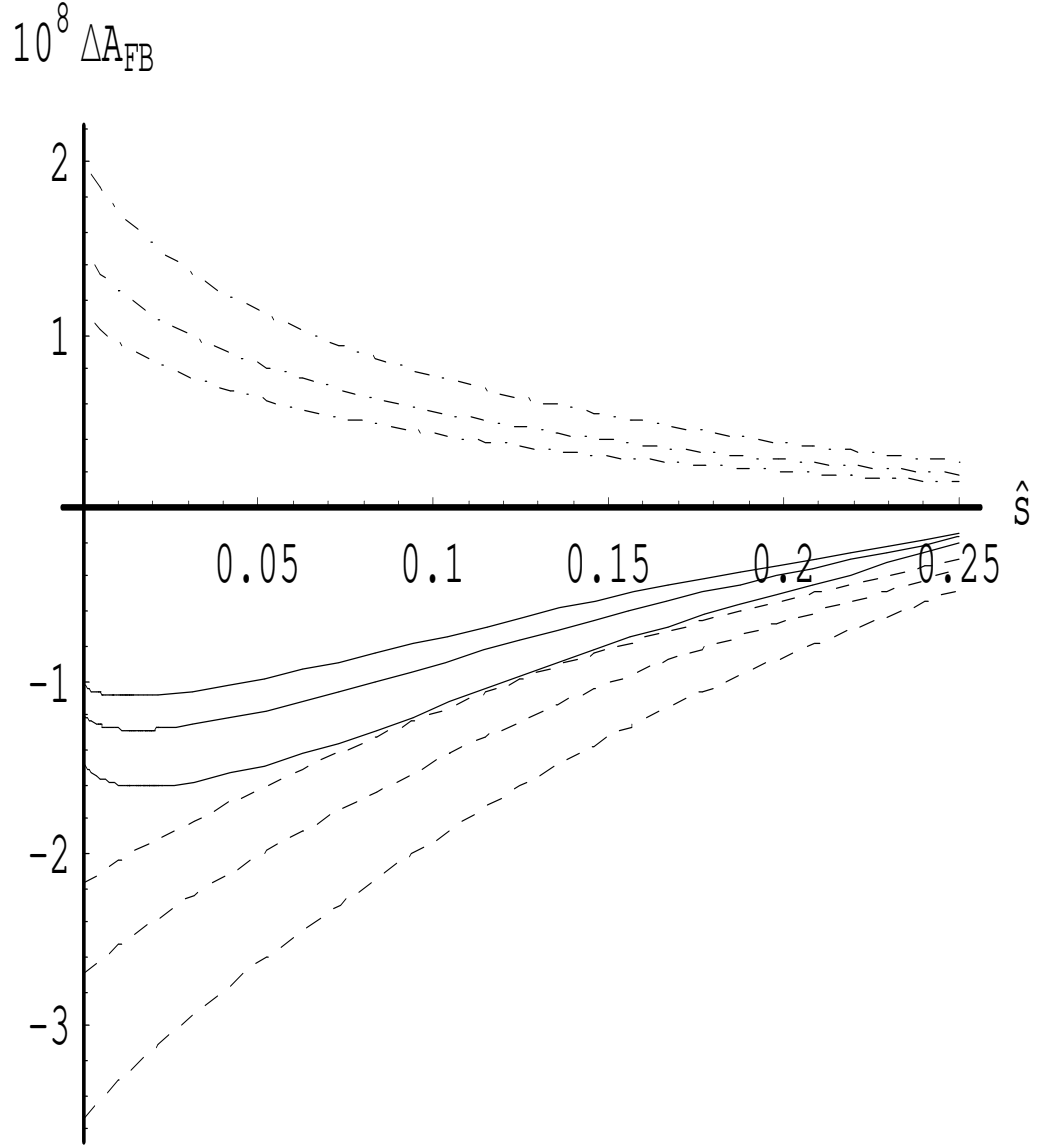}
\end{center}
\vspace{-20cm}
\caption[]{
The contribution of finite bremsstrahlung corrections to the unnormalized forward-backward
asymmetry. Dashed-dot lines show the contribution from interference of matrix elements
of operators $O_8$ and $O_{10}$  ($\mu$ =2.5 GeV (uppermost curve), 5 GeV (middle curve),
and 10 GeV  (lower curve)), dashed lines show the contribution from
interference of matrix elements of  operators $O_{1,2}$ and $O_{10}$  ($\mu$ =2.5 GeV (lower curve),
5 GeV (middle curve),  and 10 GeV (uppermost curve) and
), solid  lines show the sum of finite bremsstrahlung contributions ($\mu$ =2.5 GeV
(lower curve), 5 GeV (middle curve),  and 10 GeV (uppermost curve)), $m_c/m_b$=0.29.}
\end{figure}
\newpage
\begin{figure}
\begin{center}
\epsfxsize=18cm
\epsfysize=18cm
\leavevmode
\vspace{15cm}
\mbox{\hskip 5cm}
\epsfbox[10 7 673 377]{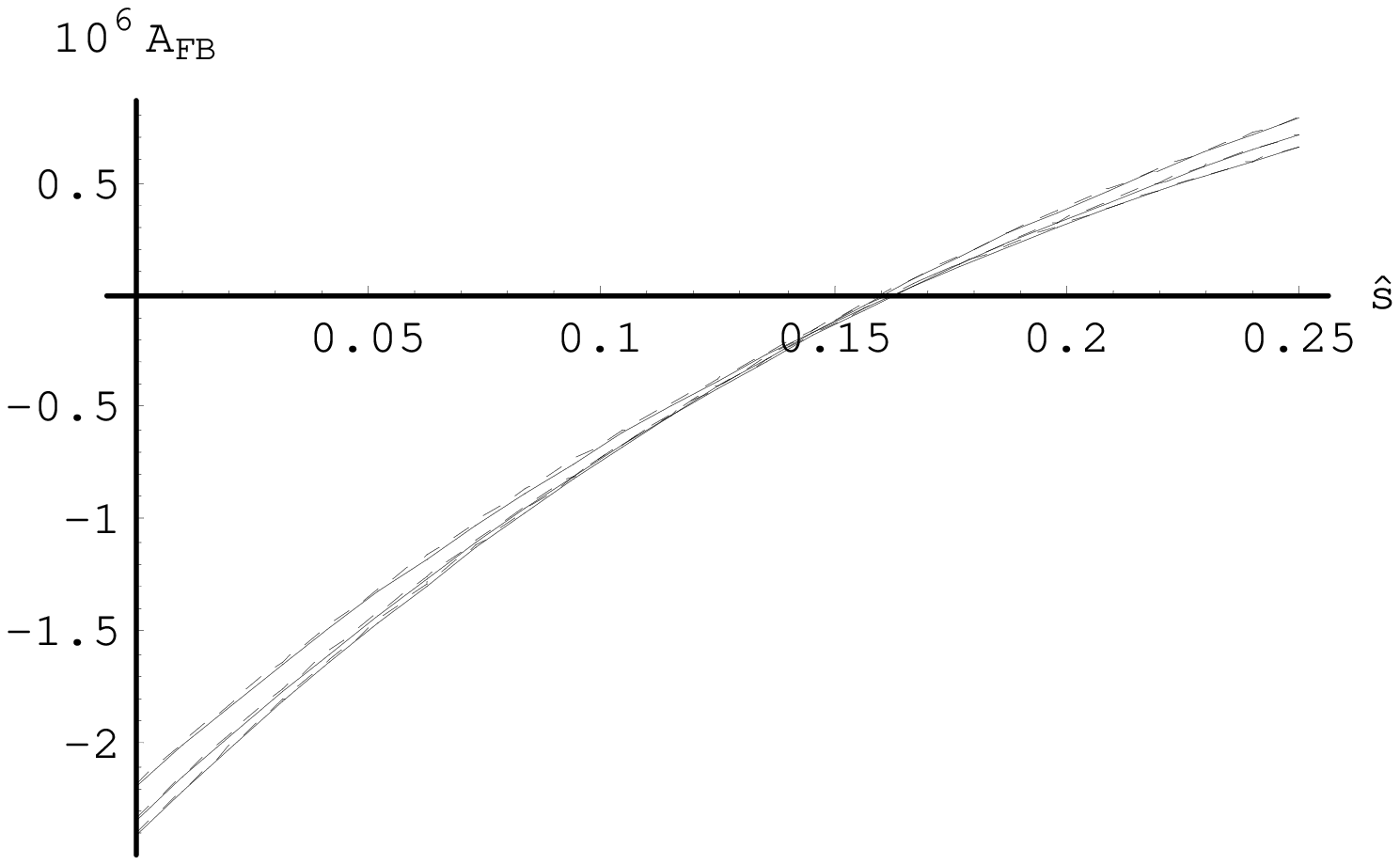}
\end{center}
\vspace{-16cm}
\caption[]{
The solid curves show $\hat{s}$ and $\mu$ (for $\mu$=2.5, 5 and 10 GeV) dependence of
unnormalized forward-backward asymmetry in NNLL approximation including finite
bremsstrahlung corrections while dashed lines show the
corresponding results  without new corrections.}
\end{figure}
\newpage
\begin{figure}
\begin{center}
\epsfxsize=18cm
\epsfysize=18cm
\leavevmode
\vspace{15cm}
\mbox{\hskip 5cm}
\epsfbox[10 7 673 377]{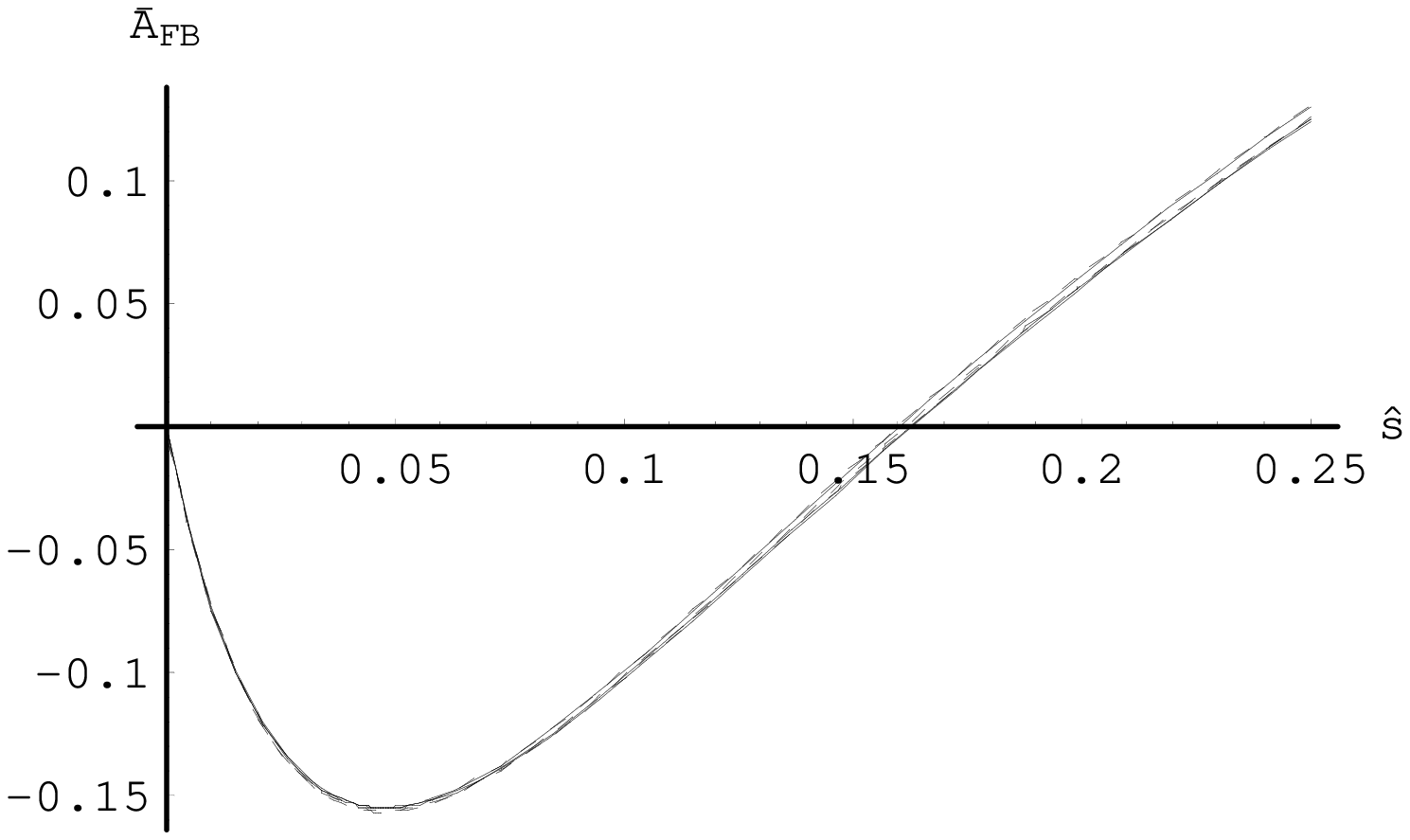}
\end{center}
\vspace{-16cm}
\caption[]{
The solid curves show $\hat{s}$ and $\mu$ (for $\mu$=2.5, 5 and 10 GeV) dependence of
normalized forward-backward asymmetry in NNLL approximation including finite
bremsstrahlung corrections while dashed lines show the
corresponding results  without new corrections.}
\end{figure}
%
\end{document}